\documentclass[prd,superscriptaddress,amsfonts,amssymb,amsmath,showpacs,onecolumn]{revtex4-2}
\usepackage{amsfonts}
\usepackage{latexsym}
\usepackage{graphicx}
\usepackage{amsmath}
\usepackage{palatino}
\usepackage{textcomp}
\usepackage{float}
\usepackage[utf8]{inputenc}
\usepackage{booktabs}
\usepackage{multirow}
\usepackage{hyperref}
\usepackage{amsmath}
\usepackage{orcidlink}
\usepackage[caption=false]{subfig}
\usepackage{commath}
\begin{document}
\title{Hamilton-Jacobi formalism for k-inflation}
\author{Rong-Jia Yang \footnote{Corresponding author}}
\email{yangrongjia@tsinghua.org.cn}
\affiliation{College of Physics Science and Technology, Hebei University, Baoding 071002, China}
\affiliation{Hebei Key Lab of Optic-Electronic Information and Materials, Hebei University, Baoding 071002, China}
\affiliation{National-Local Joint Engineering Laboratory of New Energy Photoelectric Devices, Hebei University, Baoding 071002, China}
\affiliation{Key Laboratory of High-pricision Computation and Application of Quantum Field Theory of Hebei Province, Hebei University, Baoding 071002, China}

\author{Ming Liu}
\affiliation{College of Physics Science and Technology, Hebei University, Baoding 071002, China}

\begin{abstract}
We propose a type of k-inflation under the Hamilton-Jacobi approach. We calculate various observables such as the scalar power spectrum, the tensor-to-scalar ratio, the scalar spectra index for the case where the Hubble parameter is a power-law function of k-field. The model's parameters are constrained with Planck data and the concrete form of the potential is presented. The results show that the model can be in good agreement with observations.
\end{abstract}

\maketitle

\section{Introduction}
The inflationary paradigm offers the attractive possibility of resolving some of the puzzles of standard hot big bang cosmology and also provides a mechanism for producing primary density perturbations \cite{Guth:1980zm, Linde:1981mu, Baumann:2009ds}. The mechanism that triggers the inflationary epoch is one of the most outstanding issues in contemporary cosmology. Theoretical models explaining this early accelerating expansion of the universe have exploded in recent years. Most successful inflationary scenarios are described by a single scalar field (the ``inflaton") which slowly rolls down its potential. K-inflation, described by a non-standard kinetic term, has been
one of the popular models to drive an inflationary evolution \cite{Armendariz-Picon:1999hyi}. Scalar fields with non-canonical kinetic terms arise commonly in supergravity and superstring theories and satisfy in a more natural way the slow-roll conditions of inflation. In k-inflation, since there are many plausible models, including different ways to achieve compactification, the non-canonical terms can vary significantly. Because of the non-canonical terms, the tensor-to-scalar ratio is significantly reduced in k-inflation \cite{Barenboim:2007ii,Franche:2009gk,Unnikrishnan:2012zu}. Up to now, k-inflation has been studied extensively, see for example \cite{Armendariz-Picon:1999hyi, Garriga:1999vw, Odintsov:2021lum, Granda:2021xyc, Pareek:2021lxz, Martin:2013uma, Mikura:2021ldx, Mohammadi:2019qeu, Mohammadi:2018wfk, Barenboim:2007ii, Lola:2020lvk, Shumaylov:2021qje, Feng:2014pta, Franche:2009gk, Unnikrishnan:2012zu, Gwyn:2012ey, Easson_2013, Zhang:2014dja, Rezazadeh:2014fwa, Cespedes:2015jga, Li_2012, Kamenshchik:2018sig}.

In addition to the slow-rolling approximation, Hamilton-Jacobi is another approach for studying inflation \cite{Salopek:1990jq,Muslimov:1990be,Lidsey:1991zp,Lidsey:1995np}. In this method, instead of introducing a potential the Hubble parameter is introduced as
a function of the scalar field. By doing so, the parameters of model are derived in terms of the Hubble parameter and its
first derivative. This method has been extensively applied to study various inflation models, see for example \cite{Liddle:1994dx,Kinney:1997ne,Guo:2003zf,Aghamohammadi:2014aca,Sheikhahmadi:2016wyz,Videla:2016ypa,Sayar:2017pam}. Up to our
knowledge, there is no work considering k-inflation by using this formalism. It maybe interesting to
investigate k-inflation using this powerful formalism. We will show that all of the main related parameters could easily be derived and can be determined the free parameters by using the latest observational data.

The Letter has been organised as follows: In Sec. II, we will discuss the general framework, such as k-essence cosmology, Hamilton-Jacobi formalism, attractor behavior, and cosmological perturbations. In Sec. III, as an application, we will consider a typical case where the Hubble parameter is a power-law function of k-field and discuss the observational constraints on the model. Finally, we will briefly summarize the results in section IV.

\section{General framework}
In this section, we will briefly review k-essence cosmology, Hamilton-Jacobi formalism, attractor behavior, and cosmological perturbations. We will give some fundamental equations needed in calculations.
\subsection{K-essence cosmology}
K-inflation, characterized by a scalar field with non-canonical kinetic terms, minimally coupled with gravity is described by \cite{Armendariz-Picon:2000nqq, Armendariz-Picon:2000ulo, Malquarti:2003nn}
\begin{eqnarray}
\label{1}
S=\int d^4 x\sqrt{-g}\left[-\frac{m_{\rm{p}}^{2}}{16\pi} R+\mathcal{L}_{\phi}\left(\phi, X\right)\right],
\end{eqnarray}
in which $m_{\rm{p}}$ is the Planck mass, $R$ is the Ricci scalar, $\phi$ stands for
the scalar field with kinetic terms $X=\nabla_{\mu} \phi \nabla^{\mu} \phi/2$, and $\mathcal{L}_{\phi}$ is the Lagrangian density of the scalar field. Here we take units $c=\hbar=1$. Variation comes to the field equations
\begin{equation}
\frac{\partial \mathcal{L}}{\partial \phi}-\left(\frac{1}{\sqrt{-g}}\right) \partial_\mu\left(\sqrt{-g} \frac{\partial \mathcal{L}}{\partial\left(\partial_\mu \phi\right)}\right)=0,
\end{equation}
and the components of the energy momentum tensor for scalar field in a homogeneous and isotropic universe give the pressure and the energy density, respectively
\begin{eqnarray}
\label{2}
p_{\phi}=\mathcal{L}_{\phi},~~~~~\rho_{\phi}=2X\mathcal{L}_{\phi,X}-\mathcal{L_\phi},
\end{eqnarray}
where the comma denotes the partial derivative with respect to $X$. The speed of sound for perturbations is defined as
\begin{eqnarray}
\label{ss}
c^2_{\rm{s}}=\frac{p_{,X}}{\rho_{,X}}=\frac{\mathcal{L}_{\phi,X}}{2X\mathcal{L}_{\phi,XX}+\mathcal{L}_{\phi,X}}.
\end{eqnarray}
Usually a general form of $\mathcal{L}_{\phi}$ was considered in literatures \cite{Armendariz-Picon:1999hyi, Li_2012,Mukhanov:2005bu}, however, the disadvantage of doing so is that it is not possible to obtain specific numerical values for comparison with astronomical observations. In order to facilitate comparison with observations, here we will consider a concrete form of k-inflation. Structuring a Lagrangian density with non-canonical kinetic terms, the usual approach is to add or multiply the kinetic term in the canonical Lagrangian density with the power of kinetic term: $X^{1/2}$, $X$, $X^{3/2}$, and so on. Here we consider a simple case: $(2X)^{1/2}\cdot(2X)$, where the introduced factor 2 is for the convenience of subsequent calculations. Namely, we focus on the following non-canonical Lagrangian density
\begin{eqnarray}
\label{4}
\mathcal{L}_{\phi}=p_\phi=\frac{\alpha\left(2X\right)^{\frac{3}{2}}}{m^2_{\rm{p}}}-V(\phi),
\end{eqnarray}
where $\alpha$ is a dimensionless constant. It has been shown in \cite{Bruneton:2007si} that the theory is well-defined if $\mathcal{L}_{\phi,X}> 0$ and $2X\mathcal{L}_{\phi,XX}+\mathcal{L}_{\phi,X}>0$, implying  $\alpha> 0$. From Eqs. (\ref{2}) and (\ref{4}), the corresponding energy density can be gotten as
\begin{eqnarray}
\label{6}
\rho_\phi=\frac{2\alpha\left(2X\right)^{\frac{3}{2}}}{m^2_{\rm{p}}}+V(\phi).
\end{eqnarray}
From Eq. (\ref{ss}), the speed of sound reads: $c^2_{\rm{s}}=1/2$.

Based on recent observations, the Universe is homogeneous, isotropic, and spatially flat. Such a geometry  is described by the Friedmann-Lemaitre-Robertson-Walker (FLRW) metric
\begin{eqnarray}
\label{7}
d s^{2}=d t^{2}-a^{2}(t)\left[d r^{2}+r^{2}\left(d \theta^{2}+\sin ^{2} \theta d \varphi^{2}\right)\right].
\end{eqnarray}
In this spacetime, $X=\dot{\phi^2}/2$, the Friedmann equations for k-inflation (\ref{4}) are given by
\begin{eqnarray}
\label{8}
H^{2}(\phi)=\frac{8\pi}{3m^{2}_{\rm{p}}}\rho_\phi=\frac{8\pi}{3 m^2_{\rm{p}}}V(\phi)+\frac{16\pi\alpha}{3m^4_{\rm{p}}}\dot{\phi}^3
\end{eqnarray}
\begin{eqnarray}
\label{9}
\dot{H}(\phi)=-\frac{4\pi}{m^{2}_{\rm{p}}}(\rho_\phi+p_\phi)=-\frac{12 \pi \alpha \dot{\phi}^3}{m^4_{\rm{p}}}.
\end{eqnarray}
The field equation for the k-field comes from Eq.(\ref{1}) is
\begin{eqnarray}
\label{3}
(\mathcal{L}_{\phi,X}+2X \mathcal{L}_{\phi,XX})\ddot{\phi}+3\mathcal{L}_{\phi,X} H\dot{\phi}+\mathcal{L}_{\phi,\phi}=0.
\end{eqnarray}
This equation can also be obtained from the conservation equation.

\subsection{Hamilton-Jacobi Formalism}
In Hamilton-Jacobi Formalism \cite{Salopek:1990jq,Muslimov:1990be,Lidsey:1991zp,Lidsey:1995np}, the Hubble parameter is assumed as a function of the scalar field, $H:=H(\phi)$. Therefore the time variable of $H$ can be rewritten as $\dot{H}=\dot{\phi} H^{\prime}$ with the prime denoting derivative with respect to the scalar field. Using Eq. \eqref{9}, the time derivative of the scalar field could be obtained in terms of scalar field as follows
\begin{eqnarray}
\label{dotphi}
\dot{\phi}=m^2_{\rm{p}}\sqrt{-\frac{H^{\prime}}{12\pi\alpha}}.
\end{eqnarray}
Substituting Eq. (\ref{dotphi}) into Friedmann equation (\ref{8}), we arrive at
\begin{eqnarray}
\label{hj}
H^2 + \frac{2 m^2_{\rm{p}}}{9\sqrt{3\pi\alpha}}\left(-H^{\prime}\right)^{\frac{3}{2}} - \frac{8 \pi V}{3 m^2_{\rm{p}}}=0.
\end{eqnarray}
Equation (\ref{hj}) are known as the Hamilton-Jacobi equation. The potential of the model is easily obtained as a function of the scalar field from Eq. \eqref{8}
\begin{eqnarray}
\label{potential}
V(\phi)=\frac{3 m^2_{\rm{p}}}{8\pi}H^2-\frac{m^4_{\rm{p}}}{4\sqrt{\alpha}}\left(-\frac{H^{\prime}}{3\pi}\right)^{\frac{3}{2}}.
\end{eqnarray}
The exact form of the potential is unknown and can only be assumed. Unlike the ordinary single scalar field inflationary theory, the slow-roll parameters are defined as \cite{Panotopoulos:2007ky}
\begin{eqnarray}
\label{slow}
\epsilon(\phi)=2 c_{s} m^2_{\rm{p}} \left(\frac{H^{\prime}}{H}\right)^2,~~~~\eta(\phi)=2 c_{s} m^2_{\rm{p}}\frac{ H^{\prime \prime}}{H}.
\end{eqnarray}
Considering slow roll approximation, the universe undergoes a quasi-de Sitter expansion during inflation, the slow-roll parameters must be much smaller than unity: $\epsilon, |\eta|\ll 1$. When $\ddot{a}$ vanishes, or equivalently the slow-roll
parameter $\epsilon$ arrives at unity, inflation ends. Hence we have at the end of inflation
\begin{eqnarray}
\label{slowend}
H=\sqrt{2c_{s}}m_{\rm{p}}H^{\prime}.
\end{eqnarray}
The parameter describing the amount of expansion, known as the number of e-folds, is described as
\begin{eqnarray}
\label{efold}
N \equiv \int_{t_{\rm{i}}}^{t_{\rm{e}}} H d t=\int_{\phi_{\rm{i}}}^{\phi_{\rm{e}}} \frac{H(\phi)}{\dot{\phi}} d \phi,
\end{eqnarray}
where the subscript ``i" and ``e", respectively, denote the beginning and the end of inflation. In the Hamilton-Jacobi approach,
it seems that the main parameters of model could be derived more easily than the slow-rolling approach, with less assumptions.

\subsection{Attractor behavior}
Following Ref. \cite{liddle2000cosmological} and assuming a homogeneous perturbation $\delta H$ to a solution $H(\phi)$, it is easy to consider whether all possible trajectories or solutions converge to a common attractor solution by using the Hamilton-Jacobi method. If the perturbation $\delta H$ becomes small by passing time, then the attractor condition is satisfied. Inserting $H(\phi) + \delta H(\phi)$ into Eq. (\ref{hj}) and linearizing, we obtain
\begin{eqnarray}
\label{attra}
&&H^2+2H\delta H+\frac{2 m^2_{\rm{p}}\left(-H^{\prime}\right)^{\frac{3}{2}}}{9\sqrt{3\pi\alpha}}-\frac{ m^2_{\rm{p}}(-H^{\prime})^{1/2}\delta H'}{3\sqrt{3\pi\alpha}}- \frac{8 \pi V}{3 m^2_{\rm{p}}}+\mathcal{O}(\delta H^2)=0.
\end{eqnarray}
Thinking of Eq. \eqref{hj}, we have
\begin{eqnarray}
\label{attra}
\frac{ m^2_{\rm{p}}}{3\sqrt{3\pi\alpha}} (-H^{\prime})^{1/2}\delta H'(\phi)-2H(\phi)\delta H(\phi)\simeq 0,
\end{eqnarray}
Solving this equation, yields
\begin{eqnarray}
\label{solattra}
\delta H(\phi)=\delta H(\phi_{\rm{i}})\exp\left[\frac{6\sqrt{3\pi\alpha}}{m^2_{\rm{p}}}
\int^{\phi}_{\phi_{\rm{i}}}\frac{H(\phi)}{(-H^{\prime})^{1/2}(\phi)}d\phi\right],
\end{eqnarray}
where $\delta H(\phi_{\rm{i}})$ is the initial value of perturbation. Given $H(\phi)$, we can analyse the behavior of perturbation $\delta H(\phi)$. So it
easy to consider the attractive behavior of solutions by using the Hamilton-Jacobi method.

Using (\ref{dotphi}) and (\ref{efold}), we find the following simple expression for describing the decay of perturbations
\begin{eqnarray}
\label{sola}
\delta H(\phi)=\delta H(\phi_{\rm{i}})\exp(-3N),
\end{eqnarray}
which signifies an exponentially rapid approach to the inflationary attractor solution. Remarkably the expression (\ref{sola}) does not depend on the free
parameter which characterizes our model (\ref{4}), meaning that the homogeneous perturbations in the inflationary model considered here decay in precisely the
same manner as they do for canonical scalars \cite{Salopek:1990jq}.

\subsection{Cosmological perturbations}
Consider linearized scalar and tensor perturbations to the spatially flat FLRW metric, which are described by the line element \cite{Mukhanov:1990me, Kodama:1984ziu, Bardeen:1980kt}
\begin{equation}
\mathrm{d} s^2=(1+2 A) \mathrm{d} t^2-2 a(t)\left(\partial_i B\right) \mathrm{d} t \mathrm{~d} x^i-a^2(t)\left[(1-2 \psi) \delta_{i j}+2\left(\partial_i \partial_j E\right)+h_{i j}\right] \mathrm{d} x^i \mathrm{~d} x^j,
\end{equation}
where $A$, $B$, $E$ and $\psi$ depict the scalar degree of metric perturbations while $h_{ij}$ describes tensor perturbations. The curvature perturbation $\mathcal{R}$ on the uniform field slicing is defined as a gauge invariant combination of scalar field perturbation $\delta\phi$ and the metric perturbation $\psi$
\begin{equation}
\label{R}
\mathcal{R} \equiv \psi+\left(\frac{H}{\dot{\phi}}\right) \delta \phi.
\end{equation}
From the equation governing the evolution of perturbations in the scalar field and from the linearized Einstein's equation $\delta G_{\mu\nu}=k\delta T_{\mu\nu}$, equation (\ref{R}) turns out to be
\begin{equation}
\label{rk}
\mathcal{\ddot{R}}_k+2\left(\frac{\dot{z}}{z}\right) \mathcal{\dot{R}}_k+c_s^2 k^2 \mathcal{R}_k=0,
\end{equation}
where the dot denotes derivative with respect to conformal time, $\eta=\int dt/a(t)$ and $z$ is given by
\begin{equation}
z \equiv \frac{a\left(\rho_\phi+p_\phi\right)^{1 / 2}}{c_s H}.
\end{equation}
Rewriting Eq. (\ref{rk}) in terms of the Mukhanov-Sasaki variable $u_k\equiv z \mathcal{R_k}$, one has
\begin{equation}
\ddot{u}_k+\left(c_s^2 k^2-\frac{\ddot{z}}{z}\right) u_k=0.
\end{equation}
Similarly, the equation for the tensor perturbations is
\begin{equation}
\ddot{\upsilon}_k+\left(k^2-\frac{\ddot{a}}{a}\right) \upsilon_k=0,
\end{equation}
where $\upsilon_k\equiv h/a$ with $h$ the amplitude of the tensor perturbation. The power spectrum of scalar perturbations is given by
\begin{equation}
\mathcal{P}_S(k) \equiv\left(\frac{k^3}{2 \pi^2}\right)\left|\mathcal{R}_k\right|^2=\left(\frac{k^3}{2 \pi^2}\right)\left(\frac{\left|u_k\right|}{z}\right)^2.
\end{equation}
And the tensor power spectrum defined as
\begin{equation}
\mathcal{P}_T(k) \equiv 2\left(\frac{k^3}{2 \pi^2}\right)\left|h_k\right|^2=2\left(\frac{k^3}{2 \pi^2}\right)\left(\frac{\left|v_k\right|}{a}\right)^2.
\end{equation}
The scalar and tensor perturbations is evaluated at Hubble radius crossing $c_s k=aH$ during inflation. Following \cite{Garriga:1999vw}, the scalar and tensor power spectrum in the slow roll limit can be obtained, respectively, as
\begin{eqnarray}
\label{ps}
\mathcal{P}_S(k)=\left[\frac{H^2}{2 \pi \left[c_s\left(\rho_\phi+p_\phi\right)\right]^{1/2}}\right]^2,
\end{eqnarray}
and
\begin{equation}
\label{pt}
\mathcal{P}_T(k)=\frac{16}{\pi}\left(\frac{H}{m_{\rm{p}}}\right)^2.
\end{equation}
We can constrain the model's parameters with observational data about the scalar power spectrum $\mathcal{P}_S(k)$.

\section{Application}
Up to now, we have derived basic equations and some crude results. In this section, we will consider a typical function for $ H(\phi)$ parameter in terms of the scalar field to get more specific results. Assuming that the Hubble parameter is a power-law function of k-field
\begin{eqnarray}
\label{pow}
H(\phi)=\mathcal{H}_{1} \left(\frac{\phi}{m_{\rm{p}}}\right)^{n},
\end{eqnarray}
where $n$ and $\mathcal{H}_{1}$ are constant. For simplicity, we let $\mathcal{H}_{1}=\beta m_{\rm{p}}$ with $\beta$ a dimensionless constant. From Eq. \eqref{dotphi}, we can express the time derivative of the scalar field as
\begin{eqnarray}
\label{17}
\dot{\phi}=m^2_{\rm{p}}\sqrt{-\frac{n\beta}{12\pi\alpha}}\left(\frac{\phi}{m_{\rm{p}}}\right)^{\frac{n-1}{2}}.
\end{eqnarray}
The general form for potential can be derived from Eq. \eqref{potential}
\begin{eqnarray}
\label{V}
V(\phi)=\frac{3m^4_{\rm{p}}\beta^2}{8\pi}\left(\frac{\phi}{m_{\rm{p}}}\right)^{2n}
+\frac{m^4_{\rm{p}}}{36\pi}\sqrt{-\frac{3n^3\beta^3}{\pi\alpha}}\left(\frac{\phi}{m_{\rm{p}}}\right)^{\frac{3(n-1)}{2}}.
\end{eqnarray}
The condition of having a real potential is
\begin{eqnarray}
\label{311}
\left(\frac{\phi}{m_{\rm{p}}}\right)^{\frac{n+3}{2}}\geq-\frac{2}{27}\sqrt{-\frac{3n^3}{\pi^3\alpha\beta}}\left(\frac{\phi}{m_{\rm{p}}}\right)^{\frac{3(n-1)}{2}},
\end{eqnarray}
which dependents on the parameters $n$ and $\alpha$. When the acceleration $\ddot{a}$ vanishes, inflation ends. From Eq. \eqref{slowend}, implying that the scalar field at the end of inflation could fulfil
\begin{eqnarray}
\sqrt{2 c_{s}}{\mathcal{H}_{1} n \left(\frac{\phi }{m_{\rm{p}}}\right)^{n-1}}=\mathcal{H}_{1} \left(\frac{\phi }{m_{\rm{p}}}\right)^n.
\end{eqnarray}
Solving this equation, yields
\begin{eqnarray}
\label{18}
\phi_{\rm{e}}=2^{\frac{1}{4}}m_{\rm{p}}n.
\end{eqnarray}
Using the number of $e$-folds relation \eqref{efold}, the scalar field at the begin of inflation could be obtained as follows
\begin{eqnarray}
\label{181}
\phi_{\rm{i}}=m_{\rm{p}}2^{-\frac{4}{n+3}} 3^{-\frac{2}{n+3}} \left[-12\times 2^{\frac{n+3}{8}} (-n)^{\frac{3+n}{2}}+(n+3) N \sqrt{-\frac{3n}{\pi \alpha \beta}}\right]^{\frac{2}{n+3}}.
\end{eqnarray}
The initial value of inflaton in Eq. \eqref{181} seems to depend on the model parameters. Actually, they are constants, we can phenomenologically take some special values for them, but in doing so, the model under considering may not match the observations. Here, we take another approach, treating them as undetermined constants and use observational data to determine them in subsequent calculations, as done in \cite{Sheikhahmadi:2016wyz,Aghamohammadi:2014aca}. With Friedmann equation (\ref{9}) and Hubble parameter (\ref{pow}), the scalar and tensor power spectrums in the slow roll regime are determined from Eqs. (\ref{ps}) and (\ref{pt}) to be
\begin{eqnarray}
\label{ps1}
\mathcal{P}_S=2^{\frac{-3-17n}{2(n+3)}}3^{\frac{-9n-3}{2(n+3)}}\pi^{-1/2}\alpha^{1/2}\beta^{5/2}(-n)^{-3/2}\left[-12\times2^{\frac{n+3}{8}}(-n)^{\frac{n+3}{2}}+N(n+3)\sqrt{\frac{-3n}{\pi \alpha\beta}}\right]^{\frac{5n+3}{n+3}},
\end{eqnarray}
and
\begin{eqnarray}
\label{pt1}
\mathcal{P}_T=2^{\frac{-4n+12}{n+3}}3^{\frac{-4n}{n+3}}\pi^{-1}\beta^{2}\left[-12\times2^{\frac{n+3}{8}}(-n)^{\frac{n+3}{2}}+N(n+3)\sqrt{\frac{-3n}{\pi \alpha\beta}}\right]^{\frac{4n}{n+3}}.
\end{eqnarray}
\begin{table}
\centering
\renewcommand\arraystretch{4}
\setlength{\tabcolsep}{2mm}{
\begin{tabular}{c|c c}
\hline
  & $\alpha$   & $\beta$   \\
\hline
$N=55$ & $2.37\times 10^{9}$  & $1.89\times 10^{-6}$   \\ \cline{1-3}
$N=60$ & $3.43\times10^{9}$ & $1.82\times10^{-6}$  \\ \cline{1-3}
$N=65$ & $4.70\times10^{9}$ & $1.76\times10^{-6}$  \\
\hline
\end{tabular}}
\caption{The values of the parameters $\alpha$ and $\beta$ with $\mathcal{P}_S\simeq 2\times 10^{-9}$ obtained from 2015 Planck data \cite{Planck:2015sxf}, $n=-0.07$, $r=0.02$ and $n_{\rm{s}}=0.9668$ predicted by 2018 Planck data \cite{Planck:2018jri}.
}
\label{ab}
\end{table}
\begin{figure}
\includegraphics[height=8cm,width=10cm]{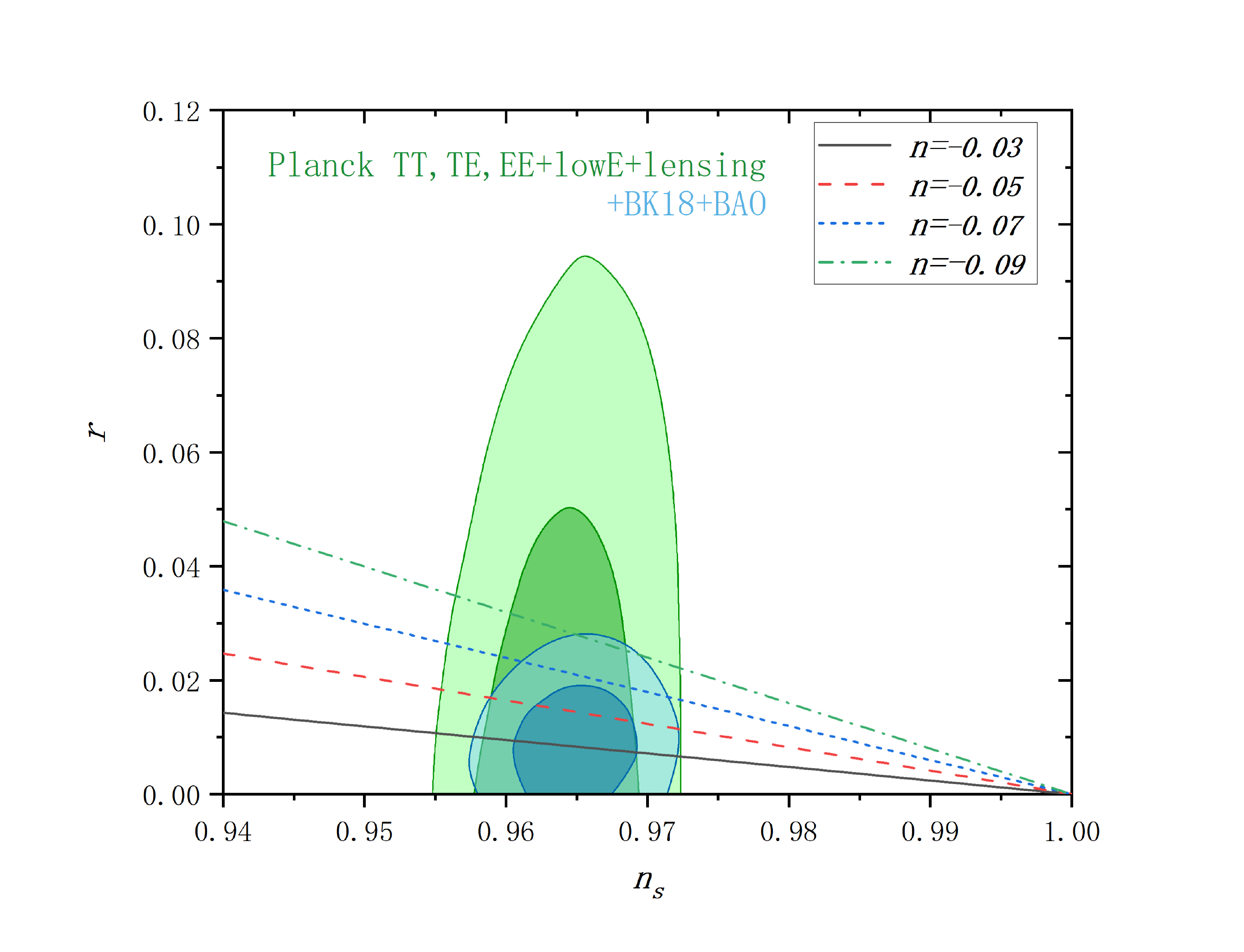}
\caption{The tensor-to-scalar ratio $r$ vs the scalar spectral index $n_{\rm{s}}$ with $n=-0.03$, $n=-0.05$, $n=-0.07$, and $n=-0.09$, respectively. Marginalized joint 68\% and 95\% C.L. regions for $r$ and $n_{\rm{s}}$ at $k=0.002$ Mpc$^{-1}$ from Planck 2018 data \cite{Planck:2018jri}, as shown in blue and green, respectively.}
\label{rns}
\end{figure}
\begin{figure}
\centering
\includegraphics[height=8cm,width=10cm]{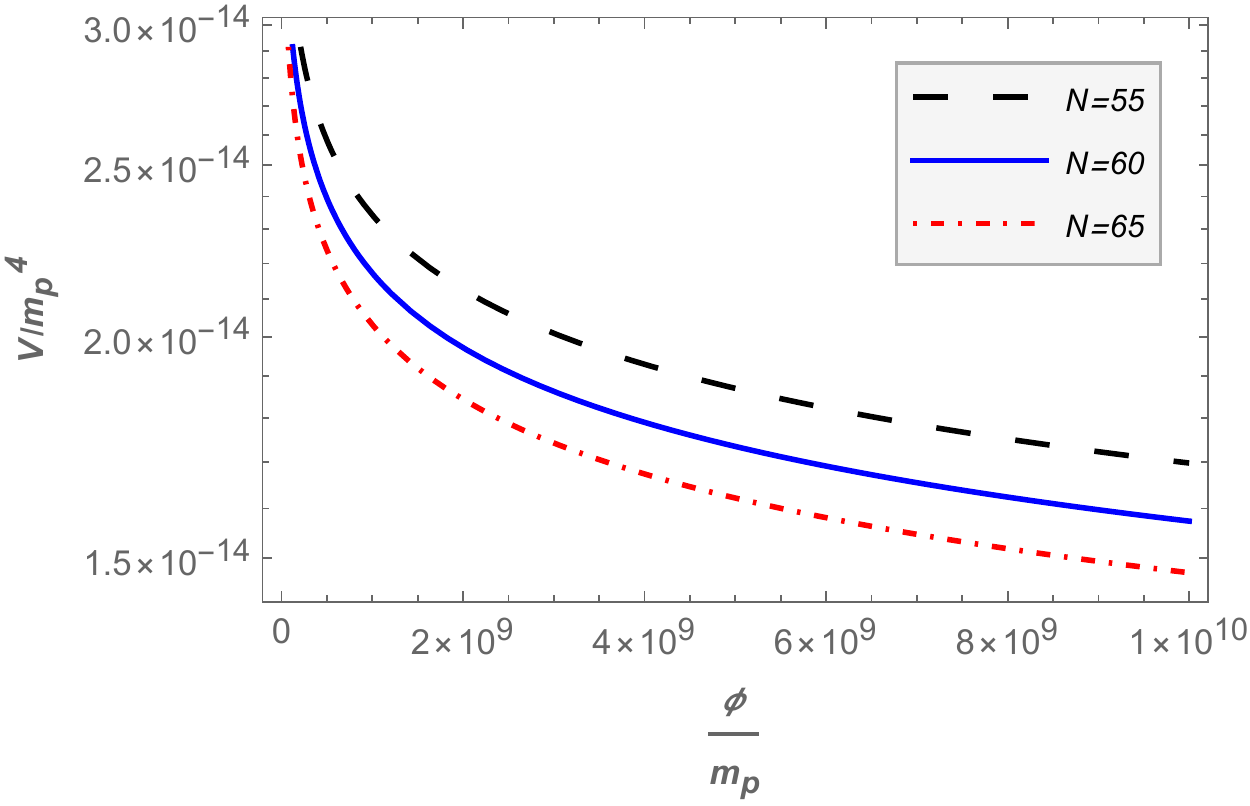}
\caption{The potential $V$ as a function of the k-field with the values of parameters in Table \ref{ab}.}
\label{fig1}
\end{figure}
Since $H$ keeps almost constant during slow roll inflation and the speed of sound is constant for our model, it turns
out that at sound horizon exit: $d/d\ln k\simeq -d/dN$. Directly coming from the amplitude of scalar and tensor perturbations, the scalar and tensor spectra indices are easily derived as
\begin{eqnarray}
\label{ns1}
n_{\rm{s}}\equiv 1+\frac{d\ln \mathcal{P}_S}{d\ln k}=1-(5n+3)\sqrt{\frac{-3n}{\pi \alpha\beta}}\left[-12\times2^{\frac{n+3}{8}}(-n)^{\frac{n+3}{2}}+N(n+3)\sqrt{\frac{-3n}{\pi \alpha\beta}}\right]^{-1},
\end{eqnarray}
and
\begin{eqnarray}
\label{nt}
n_{\rm{t}}\equiv \frac{d\ln\mathcal{P}_T }{d\ln k}=-4n\sqrt{\frac{-3n}{\pi \alpha\beta}}\left[-12\times2^{\frac{n+3}{8}}(-n)^{\frac{n+3}{2}}+N(n+3)\sqrt{\frac{-3n}{\pi \alpha\beta}}\right]^{-1}.
\end{eqnarray}
The tensor-to-scalar ratio is given by
\begin{eqnarray}
\label{r}
r\equiv\frac{\mathcal{P}_T}{\mathcal{P}_S}=-2^{\frac{9}{2}}n\sqrt{\frac{-3n}{\pi \alpha\beta}}\left[-12\times2^{\frac{n+3}{8}}(-n)^{\frac{n+3}{2}}+N(n+3)\sqrt{\frac{-3n}{\pi \alpha\beta}}\right]^{-1}.
\end{eqnarray}
From Eqs. \eqref{nt} and \eqref{r}, we observe that the consistency relation for k-inflation, $r=-8c_{\rm{s}}n_{\rm{t}}$, holds \cite{Garriga:1999vw}. Combining Eqs. (\ref{ns1}) and (\ref{r}), leads to
\begin{eqnarray}
\label{consis}
r=2^{\frac{9}{2}}\times\frac{n(n_{\rm{s}}-1)}{5n+3}.
\end{eqnarray}
Constrained from Planck TT, TE, EE+lowE+lensing+BK15+BAO \cite{Planck:2018jri}, the scalar spectral index is about $n_{\rm{s}}= 0.9668\pm0.0037$ and the tensor-to-scalar ratio $r$ is limited as: $r<0.063$. Taking $r=0.02$ and $n_{\rm{s}}=0.9668$, we get $n=-0.07$ from Eq. (\ref{consis}). Inserting these values into (\ref{ns1}), and taking $N=55$, $N=60$, and $N=65$, respectively, we can solve $\alpha$ as functions of $\beta$, and then inserting these values and $\mathcal{P}_S\simeq 2\times 10^{-9}$ \cite{Planck:2015sxf} into (\ref{ps}), we derive the values of parameter $\alpha$ and $\beta$, as listed in Table \ref{ab}.

It is interesting to compare our results with these obtained in \cite{Li_2012}, in which the spectral index $n_{\rm{s}}$ and the tensor-to-scalar ratio $r$ are computed respectively for $p(\phi,X)=K_{1+a}X^{a}-V(\phi)$ model with $V(\phi)=A\phi^n$ by using the slow-roll approximation as following
\begin{eqnarray}
\label{nsr}
&&n_{\rm{s}}=1-\frac{I(a,n)}{N},\\
\label{rsr}
&&r=\frac{8\sqrt{2a-1}n(1-n_{\rm{s}})}{n(3a-2)+2a},
\end{eqnarray}
with
\begin{eqnarray}
I(a,n)=1+\frac{(2a-1)n}{n(a-1)+2a}.
\end{eqnarray}
Obviously, Eq. (\ref{nsr}) is simpler than Eq. (\ref{ns1}), but Eq. \eqref{rsr} is more complex than Eq. \eqref{consis}; moreover, Eqs. (\ref{nsr}) and \eqref{rsr} hold only for $n>1$, while Eqs. (\ref{ns1}) and \eqref{consis} hold for real number $n$. Taking $a=3/2$ and $n_{\rm{s}}=0.9668$, we have $n=1.56$ for $N=55$, $n=1.98$ for $N=60$, and  $n=2.44$ for $N=65$ from Eq. (\ref{nsr}) obtained by using the slow-roll approximation. Inserting these values into Eq. \eqref{rsr}, we get $r=0.085$, $r=0.094$, and $r=0.101$, respectively, implying smaller $N$ is favored by observations. Taking $n_{\rm{s}}=0.9668$ and $n=-0.05$, $n=-0.07$, $n=-0.09$ respectively, we have $r=0.014$, $r=0.02$, and $r=0.027$ from Eq. \eqref{consis} derived by using Hamilton-Jacobi approach.

We plot the tensor-to-scalar ratio $r$ vs the scalar spectral index $n_{\rm{s}}$ obtained from k-inflation considered here with $n=-0.03$, $n=-0.05$, $n=-0.07$, and $n=-0.09$, respectively. Marginalized joint 68\% and 95\% C.L. regions for $r$ and $n_{\rm{s}}$ at $k=0.002$ Mpc$^{-1}$ from Planck 2018 data \cite{Planck:2018jri}, as shown in blue and green, respectively. We observe that the model is in good agreement with observations.

Inserting $n=-0.07$ and the values of $\alpha$ and $\beta$ listed in Table \ref{ab} into Eq. (\ref{potential}), we plot the scalar potential $V$ as function of the k-field
in Fig. \ref{fig1}. It is seen that the potential has a decreasing behavior and tends to zero by increasing the scalar field.

\section{conclusion}
We have proposed a type of k-inflation by using the Hamilton-Jacobi approach. After
deriving the general equation of the model, we have supposed that the Hubble parameter
could be defined as a power-law function of the k-field. The observables of the model, such as the scalar power spectrum, the tensor-to-scalar ratio, the scalar and tensor spectra indices, have been derived. The model's parameters have been constrained with Planck data and the specific form of the potential has been presented. We have compared the results here with those obtained by using the slow-roll approximation \cite{Li_2012}. We have shown with picture that the potential decreases when the k-field increases. We have plotted the tensor-to-scalar ratio $r$ vs the scalar spectral index $n_{\rm{s}}$ with some values of parameters $n$, showing that the model is in good agreement with observational data.

\begin{acknowledgments}
We thank Xinyi Zhang for helpful discussions. This study is supported in part by National Natural Science Foundation of China (Grant No. 12333008) and Hebei Provincial Natural Science Foundation of China (Grant No. A2021201034).
\end{acknowledgments}
\bibliographystyle{ieeetr}
\bibliography{reff}
\end{document}